\documentclass[11pt]{article}

\begin{document}
\large {\title{What is the topology of a Schwarzschild black hole?}
\author{EDMUNDO M. MONTE\thanks{e-mail: edmundo@fisica.ufpb.br}\\
Departamento de F\'{\i}sica, Universidade \\Federal da
Para\'{\i}ba, 58059-970, Jo\~{a}o Pessoa, Para\'{\i}ba, Brasil}

\maketitle

\begin{abstract}
We investigate the topology of Schwarzschild's black hole through the immersion of this space-time in spaces of higher dimension. Through the immersions of Kasner and Fronsdal we calculate the extension of the  Schwarzschild's black hole.
\end{abstract}

\section{Introduction}

When we want to find the gravitational field surrounding a spherically  
symmetric mass distribution at rest is obvious that this gravitational field 
should then also have spherical symmetry. Furthermore, we will require 
that the field be static. A field is said to be static if it is both time 
independent and time symmetric, i.e., unchanged by time reversal. If the 
field is merely time independent, then it is said to be stationary.

Schwarzschild space-time is the simplest relativistic model of a ''universe'' 
containing for example a single star. The star is assumed to be static and spherically 
symmetric and to be the only source of gravitation for the space-time. The 
resulting model can thus be applied to regions around any astronomical 
object that approximately fulfills these conditions. For example, in the case of 
the sun it gives a model for the solar system even better than the highly 
accurate Newtonian model. \cite{Ohanian}

The concept of black holes was considered many years
before Einstein's General Relativity in 1915. By the late
18th  century  J. Michell had reasoned, using classical physics, that light emitted radially from a
spherically symmetric body would fall back toward itself
if the radius of the body $r < 2GMc^{-2}$.  Such an object
would therefore be undetectable to the outside observer. Only with the 
Schwarzschild's solution of the Einstein's equations this radius did recognized. 
A region of space-time into which signals can enter, but from which 
no signal can ever emerge, is called a black hole. 

By  the  end  of  stellar  evolution  a  star  will  have  exhausted all its fuel having converted it to iron through
nuclear fusion. The star will cool, reducing the outwards
force, and begin to collapse under the influence of gravity.  At this point, if the star has mass greater then the
Tolman-Oppenheimer-Volkoff limit  (~ 3 solar masses)
the collapse will overcome both the electron and neutron degeneracy pressures.  No longer supportable by the
outwards force generated by the Pauli exclusion principle the inwards gravitational collapse will continue eventually forming a stellar black hole.  If the star has mass less then this limit however a white dwarf or neutron star is formed. This depends on whether it has mass less or
greater than the Chandrasekhar limit (~ 1.44 solar masses). There are others types of black holes. These
are  the  supermassive  and  primordial  black  holes;  the
former are created by the accretion of vast amounts of
matter and are usually found at the center of galaxies
whilst the latter is formed by the very large pressures present at the beginning of the universe. Today,  much astrophysical evidence has
been collected that strongly suggest the physical existence of black holes. The two main types of evidence are
gravitational lensing and intense electromagnetic flares caused by rapidly in-falling matter.  The former causes an apparent shift in position of stellar objects proportional to the magnitude of the gravitational field, whilst the latter is generated from the friction that the accreting matter. \cite{ReviewBH}

A basic result in the theory of black holes is Hawking's theorem  on the
topology of black holes,  which asserts that cross sections of the event horizon in
4-dimensional asymptotically flat stationary black hole space-times obeying the dominant energy condition are spherical (i.e., topologically $S^2$). The proof is a variational argument, showing that if a cross section has $genus = 1$ then it can be
deformed along a null hypersurface to an outer trapped surface outside of the event
horizon, which is forbidden by standard results on black holes. Hawking showed that his black hole topology result extends, 
by a similar argument, to outer apparent horizons in black hole space-times that are not necessarily stationary. A related result had been shown by Gibbons in the time-symmetric case. \cite{Gibbons}\cite{Hawking}\cite{Galloway}

The immersion problem of space-times in spaces with higher dimension 
has been required in subjects linked to minimal class of the immersion, extrinsic gravity, theory of
strings and brane-world theory. Recently we show that we can solve the Riemann tensor ambiguity problem of the space-times by embedding of the space-times in spaces of higher dimension. Furthermore,  we show that deformations of space-times of general relativity produce observable
effects that can be measured by four-dimensional observers. In the case of the FLRW cosmology, one such observable effect is shown to
be consistent with the accelerated expansion of the universe. \cite {MAJE}

Given the current interest in properties of black holes, it is of interest to
determine the topological properties of black holes  of the point of view of immersion formalism. In particular we study the topology of Schwarzschild's black hole through the immersion of this space-time in a space flat of six dimensions with different signatures. The extra dimensions are motivated by approach from \cite {MAJE} above cited. 
 
\section{The Schwarzschild's Black Hole Extension from Immersion Formalism}

We are going to return to the following problem:
Determine the
physical space outside of an approximately spherical body with
mass $M $. The physical space is  modeled through
a 4-dimensional space-time,  solution of Einstein equations, whose geometry
is  described with good approximation by 
Schwarzschild's  solution,  representing the  empty
space-time with spherical symmetry  outside of a body with spherical
mass, where  $ M =c^{2}mG^{-1}$, $c$ is the speed of light and  $G$ is the
gravitational constant. \cite{HE}

We know that in spherical coordinates $(t,r,\theta ,\phi )$, the regions
 $r=0$ and $r=2m$ are singular. When we remove the surface $r=2m$, the manifold becomes separated in two
disconnected components, one for $2m<r<\infty $ and the other for $0<r<2m $. 
The Schwarzschild black hole  $(BH_{4},g)$ is defined by:
$ BH_{4}=\{(t,r)\in \Re^{2} |\; 0<r<2m\}\;\times S^{2}$, 
where, $S^{2}$  is the sphere of radius $r$ and  the metric $g$ is
given by the usual metric of Schwarzschild. We know that $(BH_{4},g)$  may be
extensible for $r=2m$. 

Now we use the isometric immersion formalism to establish the extension of
$(BH_{4},g)$, denoted  by
$(BH'_{4},g')=(P^{2}\times S^{2},g')$, where  $P^{2}$  is the plane to be defined by immersion coordinates.

Consider two known isometric immersions of space-time from $(BH_{4},g)$ into a pseudo Euclidean 
manifold of six dimensions, with different signatures, the Kasner and Fronsdal immersions. \cite {Kasner}\cite {Fronsdal} \\
In order to determine the metric $g'$ (extension of $g$), define the  new coordinates $u$  and $v$  by:\\
- For $0<r<2m$,
\begin{equation} v=\frac{1}{4m}(\frac{-r}{2m})^{1/2}exp(\frac{r}{4m}){Y'}_{1}\;
\; \mbox{and} 
\;\;u=\frac{1}{4m}(\frac{-r}{2m})^{1/2}exp(\frac{r}{4m}){Y'}_{2},
\end{equation}
where $Y'_{1}=2(1-2m/r)^{1/2}\mbox{sinh}(t/2)$ and  $Y'_{2}=2(1-2m/r)^{1/2}\mbox{cos}h(t/2)$ 
are immersion coordinates of Fronsdal. On the other hand, 
\begin{equation}
u^{2}-v^{2}=(\frac{r}{2m}-1)exp(\frac{r}{2m})\;\Longleftrightarrow \;{%
Y^{\prime }}_{2}^{2}-{Y^{\prime }}_{1}^{2}=16m^{2}(1-\frac{2m}{r}).
\end{equation}
Now $r=r({Y^{\prime }}_{1},{Y^{\prime }}_{2})$ is implicitly defined by last
equation, while $t=t({Y^{\prime }}_{1},{Y^{\prime }}_{2})$ is
implicitly defined by 
\begin{equation}
{Y^{\prime }}_{1}/{Y^{\prime }}_{2}=tgh(\frac{t}{4m}).\label{eq:tgh}
\end{equation}
Finally, the metric $g^{\prime }$ in the new coordinates results
\begin{equation}
ds^{2} = (32m^{3}/r) exp(-r/2m)(dv^{2} -
du^{2}) - r^{2}(d\theta ^{2} +
sin^{2}\theta d\phi ^{2}).
\end{equation}

In the following section, we will prove that the topology of $(BH_{4},g)$ is
different from   the topology of $(BH'_{4},g')$. 

\section {Schwarzschild's Black Hole Topology}

Let  $(U_{\alpha},\varphi _{\alpha})$ be  a  coordinate system 
on a point $p\in M^n$ of a differentiable manifold $M^n$. Generally speaking, the
topology of a manifold $M^n$ is defined naturally 
through its open sets.  If
$A\subset M^n$,  then $A$ is an open set of $M^n$ if 
$\varphi _{\alpha}(A\cap
\varphi _{\alpha}^{-1}(U_{\alpha}))$ is an open set of $\Re^n$, $\forall \alpha$.
In other words,  the atlas of $M^n$ determines its topology.\cite{Oneill}\\

\underline{\bf Theorem:}
\vspace{3mm}

{\it The topology of Schwarzschild's black hole is given by $\Re^{2}\times S^{2}$.}
\vspace{3mm}

\underline{\bf Proof:} 
\vspace{3mm}

By  construction, $ BH_{4}=\{(t,r)\in \Re^{2} |\; 0<r<2m\}\;\times S^{2}$  and $BH'_{4}=P^{2}\times S^{2}$.
The topology of $BH_{4}$ is the Cartesian product topology of $\{(t,r)\in \Re^{2} |\; 0<r<2m\}$ by $S^{2}$, while that the topology of $BH'_{4}$ is the Cartesian product topology of $P^{2}$ by $S^2$. The topology of $S^{2}\subset \Re^{3}$
is the usual  topology induced by  the topological space $(\tau_{3},\Re^{3})$. On the other hand, the topologies of
$ \{(t,r)\in \Re^{2} |\; 0<r<2m\}\subset \Re^{2}$ and of $P^2 \subset \Re^{2}$,
respectively  $\tau_{p}$ and $\tau _{q}$,  will be induced from $(\tau_{2},\Re^{2})$.\cite{Bri}\cite{EdMa}\\ 
Since  $P^{2}$ is an extension of $\{(t,r)\in \Re^{2} |\; 0<r<2m\}$,  we may define one isometric immersion, 
\[
\psi :\;\{(t,r)\in \Re^{2} |\; 0<r<2m\}\cup \{(t,r)\in \Re^{2} |\; 2m<r<\infty\}\;\longrightarrow\;P^{2}.
\]
Therefore, for an  open set $A\subset \Re^{2}$ given by
\[
A=\{(t,r)\in \Re^{2}| \;t^{2}+(r-2m)^{2}<m^{2}\;\,\mbox{and}\;\, 0<r\},
\]
we have 
$A\cap (\{(t,r)\in \Re^{2} |\; 0<r<2m\}\cup \{(t,r)\in \Re^{2} |\; 2m<r<\infty\})=A-\{(t,r)\in \Re^{2} | \;r=2m\}$. This 
is an open set of the topological space  $\{(t,r)\in \Re^{2} |\; 0<r<2m\}\cup \{(t,r)\in \Re^{2} |\; 2m<r<\infty\}$,
composed of two connected components. Observe that open sets form a
topological basis for the semi-plane $ t-r,\;r>0$. However, we have
that 
$\psi(A\cap (\{(t,r)\in \Re^{2} |\; 0<r<2m\}\cup \{(t,r)\in \Re^{2} |\; 2m<r<\infty\}))$ is
given for an open set composed by four connected components. As the lines
$\Pi_{1}$ and $\Pi_{2}$ defined for $r=2m$ from equation $(2)$ are on $P^{2}$
we have that 
\[
[\psi (A\cap (\{(t,r)\in \Re^{2} |\; 0<r<2m\}\cup \{(t,r)\in \Re^{2} |\; 2m<r<\infty\})\cup \Pi_{1}\cup \Pi_{2}]\cap F = J ,
\]
where $F$ is an open disk on $\Re^{2}$ with center in the origin of $P^{2}$. The set 
$J$ is a  plane disk, in the new coordinates $r=r({Y'}_{1},{Y'}_{2})$
and $t=t({Y'}_{1},{Y'}_{2})$.
In this manner the topology
of $P^2$ is given by open sets of $\Re^2$. Finally 
we have that the topology of
$BH_{4}$ is equal to $\Re^{2}\times S^{2}$, clearly different
of the topology of space-time $(\{(t,r)\in \Re^{2} |\; 0<r<2m\}\cup \{(t,r)\in \Re^{2} |\; 2m<r<\infty\})$ that is $(\Re^{2}-\{(t,r)\in \Re^{2} |\;r=2m\})\times S^{2}$.$\triangle$

\section{Comments}
We prove that beginning from a Schwarzschild's black hole immersed in
an environment space flat of six dimensions and different signatures it is possible to calculate both its extension and topology. Now we can answer the question:\\
What is the topology of a Schwarzschild black hole? \\
Answer:  $\Re^{2}\times S^{2}$.

}

\end{document}